\documentclass[conference]{IEEEtran-v17}
\pdfoutput=1

\usepackage{xspace}
\usepackage[nosort]{cite}
\usepackage{url}
\usepackage[cmex10]{amsmath}
\usepackage{bbm}
\usepackage{graphicx}
\usepackage{paralist}
\usepackage{fancyref}
\usepackage[stretch=16,shrink=16,step=4]{microtype}
\usepackage{pdfsync}
\newlength{\figwidth}
\makeatletter
\def\@IEEEinterspaceratioM{0.265}
\def\@IEEEinterspaceMINratioM{0.1651}
\def\@IEEEinterspaceMAXratioM{0.38}

\def\@IEEEinterspaceratioB{0.31}
\def\@IEEEinterspaceMINratioB{0.19}
\def\@IEEEinterspaceMAXratioB{0.38}
\@IEEEtunefonts
\makeatother

 \usepackage{vmr-symbols-vecbold}
 \usepackage{standard-macros}
\safemath{\blocklength}{N}	%block-length
\safemath{\asconst}{\setO(1)} % bounded as SNR goes to infinity
\safemath{\RXant}{M} % number of receive antennas
\safemath{\rankcorr}{Q}		%rank of the correlation matrix

\safemath{\prelog}{\chi}	% pre-log
\safemath{\corrmat}{\matR}	%channel correlation matrix
\safemath{\sqrtmat}{\matP}	% sqrt of the correlation matrix

\safemath{\outvec}{\vecy} %output vector
\safemath{\outp}{y}		  %scalar output
\safemath{\inpvec}{\vecx} %input vector
\safemath{\inp}{x}		  %scalar input
\safemath{\wgnvec}{\vecw} %awgn vector
\safemath{\wgn}{w}		  %scalar awgn
\safemath{\altwgn}{z}	  %alt. scalar awgn
\safemath{\ch}{s}		 	  %scalar channel
\safemath{\chvec}{\vecs}	%channel vector
\safemath{\outvecnn}{\vecr}	%output vector (no noise)
\safemath{\chgenvec}{\vech}	%generic vector channel
\safemath{\chgenvecdir}{\widehat{\chgenvec}}	%generic vector channel direction
\safemath{\chgen}{h}		%generic component of the vector channel
\safemath{\inpprobmeas}{\mathsf{Q}}%{\setQ}	% input probability measure
\safemath{\altinpprobmeas}{\widetilde{\inpprobmeas}}
\safemath{\outprobmeas}{\mathsf{R}}%{\setR}	% output probability measure
\safemath{\chtran}{\mathsf{W}}%{W}			% channel transition probability
\safemath{\dtime}{n}			% time-slot inde
\safemath{\chmat}{\matS}		% iid channel matrix (for SIMO)

\safemath{\inpmin}{\inp\sub{min}}

\newtheorem{thm}{Theorem}
\newtheorem{lem}[thm]{Lemma}

\begin{document}
\IEEEoverridecommandlockouts

\title{Capacity Pre-Log of SIMO Correlated Block-Fading Channels\thanks{The work of Erwin Riegler was supported by the WWTF project NOWIRE.}}

\author{\IEEEauthorblockN{Wei Yang$^1$, Giuseppe Durisi$^1$, Veniamin I. Morgenshtern$^2$, Erwin Riegler$^3$
}\\
\IEEEauthorblockA{
$^1$Chalmers University of Technology, 41296 Gothenburg, Sweden\\
$^2$ETH Zurich, 8092 Zurich, Switzerland\\
$^3$Vienna University of Technology, 1040 Vienna, Austria
}}

% make the title area
\maketitle

%%%%%%%%%%%%%%%%
\begin{abstract}
We establish an upper bound on the noncoherent capacity pre-log of  temporally correlated block-fading single-input multiple-output (SIMO) channels.
The upper bound matches the lower bound recently reported in Riegler \emph{et al.} (2011), and, hence, yields a complete characterization of the SIMO noncoherent capacity pre-log, provided that the channel covariance matrix satisfies a mild technical condition.
This result allows one to determine the optimal number of receive antennas to be used to maximize the capacity pre-log for a given block-length and a given rank of the channel covariance matrix.
\end{abstract}
\IEEEpeerreviewmaketitle
%%%%%%%%%%%%%%%%%%%%%%%%%%%%%%%%%%%%%%%%%%%%%

\section{Introduction} % (fold)
\label{sec:introduction}
A crucial step in the design of wireless communication systems operating over fading channels is to determine the optimal amount of resources to be used for channel estimation.
A fruitful approach to address this problem in a fundamental fashion is to characterize the channel capacity \emph{pre-log} (i.e., the asymptotic ratio between capacity and the logarithm of the signal-to-noise ratio (SNR) as SNR goes to infinity) in the \emph{noncoherent setting} where neither transmitter nor receiver are aware of the realization of the fading process, but both know its statistics perfectly.\footnote{Capacity in the noncoherent setting is often referred to as noncoherent capacity.
In the remainder of this paper, it will be referred to simply as capacity.}
While a capacity pre-log characterization for single-input single-output (SISO) systems is available for several fading models of practical interest~\cite{marzetta99-01a,hochwald00-03a,liang04-09a,lapidoth03-10a}, the multiple-input multiple-output (MIMO) case is still largely open.

The impact of multiple antennas on the capacity pre-log has been characterized in~\cite{zheng02-02a} for the Rayleigh-fading \emph{constant block-fading} model.
According to this model, the channel stays constant over a block of \blocklength channel uses
 and changes in an independent fashion from block to block.
The approach used in~\cite{zheng02-02a} to characterize the capacity pre-log is based on an apposite change of variables, which reveals the geometry in the problem.
One interesting consequence of the analysis in~\cite{zheng02-02a} is that the SISO capacity pre-log of constant block-fading channels coincides with the single-input multiple-output (SIMO) capacity pre-log.
Hence, using multiple antennas at the receiver only does not yield a larger capacity pre-log.

A more accurate yet simple way to capture channel variations in time is to assume that the channel is correlated (but not necessarily constant) in each block, with the rank of the corresponding $\blocklength \times \blocklength$ correlation matrix given by~\rankcorr.
We shall refer to this model as \emph{correlated block-fading}.
For this channel model, the SISO capacity pre-log was determined in~\cite{liang04-09a}, whereas the MIMO case is still open.
A lower bound on the SIMO capacity pre-log was recently reported in~\cite{morgenshtern10-06a} and refined in~\cite{riegler11-08a}.
The results in~\cite{morgenshtern10-06a,riegler11-08a} are surprising, as they imply that, when $\rankcorr>1$, the SIMO pre-log can be larger than the SISO pre-log.
%\todo{Say that this result implies that less pilots need to be used.}

% section introduction (end)

\paragraph*{Contributions} % (fold)
\label{sec:contributions}
In this paper, we provide an upper bound on the SIMO capacity pre-log that matches the lower bound reported in \cite{riegler11-08a}.
Hence, the  SIMO capacity pre-log is fully characterized.
Our result allows us to establish that the optimal number of receive antennas to be used to maximize the capacity pre-log for a given block-length~\blocklength and rank~$\rankcorr<\blocklength$ of the channel correlation matrix  is    $\lceil (\blocklength-1)/(\blocklength-\rankcorr)\rceil$.
%, where $\lceil\cdot\rceil$ denotes the minimum integer larger than $(\cdot)$.
%
%\todo{Say something about what we can conclude from this characterization that we did not know before}
% Our result implies that the pre-log gets larger than the SISO case by increasing the number of receive antennas $\RXant$. And the optimal number of receive antennas to maximize the pre-log is $\lceil (\blocklength-1)/(\blocklength-\rankcorr)\rceil$, where $\lceil\cdot\rceil$ denotes the minimum integer larger than $(\cdot)$.

For the constant block-fading case, we provide an alternative and much simpler derivation of the SIMO capacity  pre-log than the one provided in~\cite{zheng02-02a}.
% Furthermore, differently from the characterization in \cite[Lem.~13]{zheng02-02a}, which holds for $\blocklength \geq \RXant +1$, our result holds for general $\blocklength \geq 2$.
Our proof is based on \emph{duality}~\cite{lapidoth03-10a} and fully exploits the geometry in the problem unveiled in~\cite{zheng02-02a}.

\paragraph*{Notation} % (fold)
\label{sec:notation}
Uppercase boldface letters denote matrices and lowercase boldface letters designate vectors.
% Uppercase sans-serif letters (e.g.,~$\mathsf{Q}$) denote probability distributions, while lowercase sans-serif letters  (e.g.,~$\mathsf{r}$) are reserved for probability density functions (pdf).
The superscripts~$\tp{}$ and~$\herm{}$ stand for transposition and Hermitian transposition, respectively.
For a matrix~$\matA \in \complexset^{m\times n}$, we write~$\veca_i$ for its $i$th column, $\tr\{\matA\}$ for its trace, and $\sigma_i(\matA)$ for its $i$th largest singular value.
%Furthermore, we denote by~$[\veca_1\cdots\veca_n]$ a matrix having as columns the vectors~$\veca_1,\ldots,\veca_n$.
% If $\setI \subset \{1,\ldots,m\}$, we use $\matA_{\setI}\in \complexset^{\abs{\setI}\times n}$ to denote the submatrix of $\matA$ retaining all rows of $\matA$ with row index in $\setI$.
 For a vector~$\veca$, $\diag\{\veca\}$ denotes the diagonal matrix that has the entries of~$\veca$ on its main diagonal and $a_i$ denotes the $i$th entry of $\veca$.
 We use a combination of superscripts and subscripts to indicate sequences of random variables or vectors.
For example, $\veca_m^n$ denotes the sequence of random vectors $\veca_m,\veca_{m+1}\ldots,\veca_n$.
We use $\abs{\setI}$ to denote the cardinality of the set $\setI$.
 We denote expectation by~$\Ex{}{\cdot}$ and use the notation~$\Ex{\vecx}{\cdot}$ or $\Ex{\inpprobmeas}{\cdot}$ to stress that expectation is taken with
respect to \vecx with probability distribution~$\inpprobmeas$.
%The notation $\log(\cdot)$ indicates the natural logarithm.
The relative entropy between two probability distributions $\inpprobmeas$ and $\outprobmeas$ is denoted by $\mathrm{D}(\inpprobmeas\|\outprobmeas)$.
For two functions~$f(x)$ and~$g(x)$, the
notation~$f(x) = \setO(g(x))$, $x\to \infty$, means that
$\lim \sup_{x\to\infty}\bigl|f(x)/g(x)\bigr|<\infty$, and
$f(x) = \landauo(g(x))$, $x\to \infty$, means that $\lim_{x\to\infty}\bigl|f(x)/g(x)\bigr|=0$.
%use $\lceil x \rceil$ to denote the minimum integer larger than $x$ and
For two random matrices $\matA$ and $\matB$, we write $\matA \stackrel{d}{=} \matB $ to indicate that $\matA$ and $\matB$ have the same distribution.
Finally, $\jpg(\veczero,\matR)$ stands for the distribution of a circularly-symmetric complex Gaussian random vector with covariance matrix $\bR$.
% subsection notation (end)
% subsection contributions (end)

\section{System Model} % (fold)
\label{sec:system_model}
%\todo{Need to replace $\sqrtmat$ with $\tp{\sqrtmat}$ and rows by column...}
We consider a Rayleigh-fading correlated block-fading SIMO channel with block-length \blocklength and \RXant receive antennas.
The main feature of the correlated block-fading model is that the fading in each component channel between the transmit antenna and each receive antenna is independent across blocks of \blocklength channel uses, but is correlated within each block, with the rank of the corresponding covariance matrix given by $\rankcorr\leq \blocklength$.
We shall also assume that the fading is independent and identically distributed (\iid) across component channels.
% Denoting by $\herm{\sqrtmat}\sqrtmat$, with $\sqrtmat \in \complexset^{\rankcorr \times \blocklength}$ and $\rank\{\sqrtmat\}=\rankcorr$, the channel covariance matrix,
The input-output~(IO) relation within a block of \blocklength channel uses can be conveniently expressed in matrix form as follows:
\begin{align}\label{eq:IO_relation}
	\matY = \matS \tp{\sqrtmat}\diag\{\vecx\} + \matW.
\end{align}
Here, $\vecx \in  \complexset^{\blocklength}$ contains the input symbols transmitted within the block.
We assume that \vecx is subject to the following average-power constraint:
\begin{equation}
\label{eq:average-power-constraint}
 \Ex{}{\vecnorm{\vecx}^2} \leq \blocklength\snr.
\end{equation}
The \emph{whitened} fading matrix \matS is of size $\RXant \times \rankcorr$ and has \iid $\jpg(0,1)$ entries.
The $\blocklength \times \rankcorr$ matrix \sqrtmat, which is deterministic and of full rank $\rankcorr\leq \blocklength$, describes the correlation structure within a block.
We shall assume that the rows of \sqrtmat have unit norm, and, hence, that the entries of the matrix $ \matS \tp{\sqrtmat}$ are identically distributed.
Finally, the $\RXant \times \blocklength$ Gaussian noise matrix~\matW has \iid  $\jpg(0,1)$ entries, and the $\RXant \times \blocklength$ matrix~\matY collects the signals from the \RXant receive antennas during  $\blocklength$ channel uses.
The model just described is of practical relevance, because it captures channel variation in time in an accurate but simple way: large \rankcorr corresponds to fast channel variation.
Furthermore,~\eqref{eq:IO_relation} models accurately the IO relation in the frequency domain of a cyclic-prefix orthogonal frequency-division multiplexing system that operates over a multipath channel with \rankcorr uncorrelated taps.
Note that, when~$\rankcorr=1$, the correlated block-fading model reduces to the constant block-fading model.

The capacity of the channel~\eqref{eq:IO_relation} is given by
\begin{equation*}
%\label{definition_capacity}
\capacity(\snr) \triangleq \frac{1}{\blocklength} \sup_{\mathsf{Q}}\mi(\vecx;\matY)
\end{equation*}
where $\mi(\vecx;\matY)$ denotes the mutual information between $\vecx$ and $\matY$ in \eqref{eq:IO_relation}, and the supremum is over all probability distributions $\mathsf{Q}$ on $\vecx$ that satisfy~\fref{eq:average-power-constraint}.
As the noise has unit variance, \snr denotes the SNR.
The capacity pre-log \prelog is defined as
\begin{equation*}	
	\prelog=\lim_{\snr \to \infty} {\capacity(\snr)}/{\log \snr}.
\end{equation*}
%
%\todo{Should recall that we consider the noncoherent setting}
% section system_model_and_known_results (end)

\section{Known Results} % (fold)
\label{sec:known_results}
In the noncoherent setting where the realizations of the fading process \matS are not known to transmitter and receiver (but \sqrtmat and the statistics of \matS are perfectly known), an analytic characterization of $\capacity(\snr)$ is not available.
As we shall review next, pre-log expressions are available for some values of \blocklength, \rankcorr, and \RXant.

For the SISO case ($\RXant=1$), Liang and Veeravalli~\cite{liang04-09a} proved that the pre-log is equal to $1-\rankcorr/\blocklength$.
This result can be interpreted as follows: channel uncertainty yields a penalty of $\rankcorr/\blocklength$ compared to the case when the channel is perfectly known to the receiver (in this case, capacity grows logarithmically with SNR and the capacity pre-log is one~\cite{biglieri98-10a}).
Alternatively, we can interpret $\rankcorr/\blocklength$ as the fraction of channel uses in which \emph{pilot symbols} need to be transmitted to learn the channel at the receiver~\cite{durisi11-08a}.
When $\rankcorr=\blocklength$, learning the channel requires to transmit pilot symbols in each channel use; hence, $\prelog=0$.
In this case, capacity  turns out to grow double-logarithmically with SNR, independently of the number of receive antennas~\cite[Thm.~4.2]{lapidoth03-10a}.

For the special case $\rankcorr=1$ (i.e., constant block-fading), the SISO capacity can actually be characterized up to a $\landauo(1)$ term~\cite{hochwald00-03a,zheng02-02a} (see~\cite{durisi11-08a} for a simple proof).
For the SIMO case, such a characterization is available only when $\blocklength\geq \RXant +1$~\cite[Lem.~13]{zheng02-02a}.
However, a pre-log characterization is available for all block-length values~\blocklength.
In particular, it follows from~\cite[Eq.~(27)]{zheng02-02a} that the SIMO capacity pre-log for the $\rankcorr=1$ case is equal to $1-1/\blocklength$, i.e., it coincides with the SISO capacity pre-log.
%that the pre-log is equal to $1-1/\blocklength$ whenever $\blocklength\geq \RXant+1$.\todo{Explain in a footnote that they provide a geometric argument for the fact that the pre-log is the same also when $\blocklength< \RXant+1$}
%This result implies that, when $\rankcorr=1$,  the SIMO pre-log is equal to the SISO pre-log.
%
This result implies that, when $\rankcorr=1$, using multiple antennas at the receiver only is not beneficial from a pre-log point of view.

This statement turns out to be no longer valid when $\rankcorr> 1$.
% Recently, it has been shown that for the general case $\rankcorr\geq 1$ this statement is no longer valid, and that adding receive antennas can actually increase the pre-log.
More precisely, the following result was recently proven in~\cite{riegler11-08a}:
\begin{thm}[{\cite[Thm.~1]{riegler11-08a}}]\label{thm:erwin}
	Suppose that $\sqrtmat$ in~\eqref{eq:IO_relation} satisfies the following
	\emph{Property (A)}: There exists a subset of indices $\setK \subset \{1,\dots,\blocklength\}$ with cardinality
	\begin{equation*}
	\abs{\setK} \define \min ( \lceil(\rankcorr \RXant -1)/(\RXant-1)\rceil, \blocklength)
	\end{equation*}
	such that every $\rankcorr$ row vectors of the submatrix of $\sqrtmat$ obtained by retaining the rows in \sqrtmat with indices in \setK are linearly independent.
	Then the pre-log of the channel~\eqref{eq:IO_relation} is lower-bounded as
	\begin{align*}
	   \prelog \geq \min \left\{\RXant \left(1-{\rankcorr}/{\blocklength}\right), 1-{1}/{\blocklength}\right\}.
	\end{align*}
\end{thm}

\fref{thm:erwin} implies that the pre-log  penalty of $\rankcorr/\blocklength$ incurred in the SISO case by not knowing the channel at the receiver can be reduced to $1/\blocklength$ by deploying multiple antennas at the receiver side, as long as the block-length is sufficiently large and \sqrtmat satisfies Property~(A).
In other words, one pilot symbol per block suffices to learn the channel at the receiver.
Intuitively, Property~(A) ensures that one can recover both $\matS$ and $\blocklength-1$ entries of \vecx from the noiseless receive signal $\matS\tp{\sqrtmat}\diag\{\vecx\}$, once one entry of \vecx is fixed~\cite{riegler11-08a}.
%
%For the special case $\rankcorr=1<\blocklength$ a more accurate capacity characterization is
% section known_results (end)
%
%
\section{A Matching Pre-Log Upper Bound} % (fold)
\label{sec:the_simo_capacity_pre_log}
The main result of this paper is the following theorem:
\begin{thm}\label{thm:our_result}
The capacity pre-log of the channel~\eqref{eq:IO_relation} is upper-bounded by
\begin{equation}
\label{eq:prelog_ub}
\prelog \leq \min \left\{ \RXant \left(1-{Q}/{N}\right), 1-{1}/{N}\right\}. %, \quad \rho \rightarrow \infty
\end{equation}
\end{thm}
\paragraph*{Remarks} % (fold)
\label{par:remarks}
\fref{thm:our_result}, combined with~\fref{thm:erwin}, yields a complete characterization of the SIMO capacity pre-log for the case when \sqrtmat satisfies Property~(A).
The SIMO capacity pre-log is given by
%
% \begin{equation}\label{eq:pre-log}
% 	   \prelog=\min \left\{ \RXant \left(1-{Q}/{N}\right), 1-{1}/{N}\right\}
% \end{equation}
%
%which is
the minimum between the number of receive antennas \RXant times the SISO capacity pre-log of a rank-$\rankcorr$ channel, and the SISO capacity pre-log of a rank-$1$ channel.
Note that the pre-log upper bound in~\eqref{eq:prelog_ub} holds independently of whether \sqrtmat satisfies Property~(A) or not.
We expect the upper bound to be loose if~Property~(A) is not satisfied.
Assume now that every $\rankcorr\times \rankcorr$ submatrix of \sqrtmat has full rank  (a condition slightly stronger than Property~(A)).
Then,~\eqref{eq:prelog_ub} implies  that the optimal number of receive antennas to be used to maximize the capacity pre-log for a given block-length~\blocklength and rank~$\rankcorr<\blocklength$ of the channel correlation matrix  is    $\lceil (\blocklength-1)/(\blocklength-\rankcorr)\rceil$.
%

% paragraph remarks (end)

\paragraph*{Outline of the proof} % (fold)
\label{par:outline_of_the_proof}
The proof consists of two parts.
We first prove that $\prelog\leq \RXant(1-Q/N)$ by generalizing to the SIMO case the approach used in~\cite[Prop.~4]{liang04-09a} to establish a tight upper bound on the SISO capacity pre-log.
Then, we prove that $\prelog\leq 1-1/\blocklength$ by showing that the capacity of a rank-$\rankcorr$ channel with $\RXant$ receive antennas can be upper-bounded by the capacity of a rank-$1$ channel with $\RXant\rankcorr$ receive antennas.
The desired result then follows  by~\cite[Eq.~(27)]{zheng02-02a}.
As the proof of~\cite[Eq.~(27)]{zheng02-02a} is rather involved, we provide an alternative, much simpler proof of this result (for the SIMO case) in \fref{sec:proof_of_lemma_rank_one}.
% paragraph outline_of_the_proof (end)

\section{Proof of~\fref{thm:our_result}} % (fold)
\label{sec:proof_of_thm:our_result}

\paragraph*{First Part: $\prelog\leq \RXant(1-\rankcorr/\blocklength)$} % (fold)
%\label{sec:first_part}
Without loss of generality, we assume that the first \rankcorr rows of~$\sqrtmat$ are linearly independent.
This can always be achieved by rearranging the columns of \matY in~\eqref{eq:IO_relation}.
We start by manipulating $I(\vecx;\matY)$ as follows (we use the notation convention introduced in \fref{sec:introduction}):
% Let $\vecy_k$ denote the $k$th column of \matY.
% Similarly, let $x_k$ denote the $k$th entry of \vecx.
% Then
%
\begin{align}\label{eq:splitting_mi}
	I(\vecx;\matY)&=I(x_1^{\blocklength};\vecy_1^{\blocklength})\notag \\
	&\stackrel{(a)}{=}I(x_1^{\blocklength};\vecy_1^{\rankcorr})+I(x_1^{\blocklength};\vecy_{\rankcorr+1}^{\blocklength}  \given \vecy_1^{\rankcorr})\notag\\
	&\stackrel{(b)}{=}I( x_1^{\rankcorr};\vecy_1^{\rankcorr})+I(x_1^{\blocklength};\vecy_{\rankcorr+1}^{\blocklength}\given \vecy_1^{\rankcorr}).
\end{align}
Here, in~(a) we used chain rule for mutual information and~(b) follows because   $\vecy_1^{\rankcorr}$ and $x_{\rankcorr+1}^{\blocklength}$ are conditionally independent given $x_1^{\rankcorr}$.
We next upper-bound each term on the right-hand side (RHS) of~\eqref{eq:splitting_mi} separately.
The assumption that the first \rankcorr rows of \sqrtmat are linearly independent implies that the first term on the RHS of~\eqref{eq:splitting_mi} grows at most double-logarithmically with SNR.
More precisely, we have that~\cite[Thm.~4.2]{lapidoth03-10a}:
\begin{align}\label{eq:first_part_first_term}
	    I(x_1^{\rankcorr};\vecy_1^{\rankcorr})\leq \log\log \snr + \landauO(1), \quad \snr \to \infty.
\end{align}
For the second term on the RHS of~\eqref{eq:splitting_mi}, we proceed as follows:
\begin{align}
\label{eq:first_part_second_term}
	&I(x_1^{\blocklength};\vecy_{\rankcorr+1}^{\blocklength} \given \vecy_1^{\rankcorr})
	=\difent(\vecy_{\rankcorr+1}^\blocklength \given \vecy_1^{\rankcorr})-\difent(\vecy_{\rankcorr+1}^\blocklength \given \vecy_1^{\rankcorr},  x_1^{\blocklength})\notag\\
	&\stackrel{(a)}{\leq} \difent(\vecy_{\rankcorr+1}^\blocklength)-\difent(\vecy_{\rankcorr+1}^\blocklength \given \vecy_1^{\rankcorr},  x_1^{\blocklength},\matS)\notag\\
	&=\difent(\vecy_{\rankcorr+1}^\blocklength) - \difent(\vecw_{\rankcorr+1}^\blocklength)\notag\\
	&\stackrel{(b)}{\leq} \sum_{k=Q+1}^\blocklength \difent(\vecy_k)+\asconst,\quad \snr \to \infty \notag\\
	&\stackrel{(c)}{\leq} \sum_{k=Q+1}^\blocklength \RXant \log\lefto(1+\Ex{}{\abs{x_k}^2}\right) + \asconst, \quad \snr \to \infty \notag\\
	&\stackrel{(d)}{\leq} \sum_{k=Q+1}^\blocklength \RXant \log(1+\blocklength\snr)+ \asconst,\quad \snr \to \infty\notag\\
	&=\RXant (\blocklength-\rankcorr)\log\snr+\asconst,\quad \snr \to \infty.
	\end{align}
Here, in (a) we used that conditioning reduces entropy; (b) follows by chain rule for differential entropy and because conditioning reduces entropy; (c) follows  because jointly proper Gaussian random vectors are entropy-maximizers for a fixed covariance matrix and because  $\Ex{}{\vecy_k\herm{\vecy_k}}=(1+\Ex{}{|x_k|^2})\matI_{\RXant}$ (recall that we assumed that the rows of \sqrtmat have unit norm); finally, in (d) we used the average-power constraint~\eqref{eq:average-power-constraint}.
The desired upper bound on the capacity pre-log follows by substituting \eqref{eq:first_part_first_term} and~\eqref{eq:first_part_second_term} into~\eqref{eq:splitting_mi}.
% sqrtmatsubsection_first_part (end)

\paragraph*{Second part: $\prelog\leq 1-1/\blocklength$} % (fold)
%\label{sec:second_part_}
%
We show that the capacity of a rank-$\rankcorr$ channel with $\RXant$ receive antennas is upper-bounded by the capacity of a rank-$1$ channel with $\rankcorr\RXant$ receive antennas.
By simple matrix manipulations, we can rewrite the IO relation~\eqref{eq:IO_relation} in the following more convenient form:
\begin{equation*}
%\label{eq:alt_IO_relation}
\matY = \sum\limits_{q=1}^{\rankcorr} \vecs_q \tp{\vecx} \diag\{\vecp_q\} + \matW.
\end{equation*}
Let now $\matW_1, \cdots,\matW_\rankcorr$ be $\RXant \times \blocklength$ independent random matrices with \iid $\jpg(0,1)$ entries.
As, by assumption, the rows of \sqrtmat have unit norm, we have that
\begin{equation*}
\matW \stackrel{d}{=} \sum\limits_{q=1}^\rankcorr \matW_q \diag\{\vecp_q\}.
\end{equation*}
Hence, we can rewrite $\matY$ as
\begin{equation*}
%\label{eq:NEW-relation}
\matY \stackrel{d}{=} \sum \limits_{q=1}^{\rankcorr} \matY_q \diag\{\vecp_q\}
\end{equation*}
where
\begin{equation*}
\matY_q\define \vecs_q\tp{\vecx}+\matW_q.
\end{equation*}
Note now that each $\matY_q$ is the output of a rank-$1$ SIMO channel with $\RXant$ receive antennas.
By observing that $\vecx$ and $\matY$ are conditionally independent given $\{\matY_1,\cdots,\matY_{\rankcorr}\}$, we conclude that, by the data-processing inequality~\cite[Sec. 2.8]{cover06-a},
\begin{equation*}
\mi(\vecx;\matY)\leq\mi\left(\vecx;\matY_1,\ldots,\matY_{\rankcorr}\right).
\end{equation*}
The claim follows by noting that the $(\rankcorr\RXant) \times \blocklength$ matrix obtained by stacking the matrices $\matY_q$ on top of each others is the output of a rank-$1$ SIMO channel with $\rankcorr\RXant$ receive antennas.
As reviewed in~\fref{sec:known_results}, the SIMO capacity pre-log for the rank-$1$ case coincides with the SISO capacity pre-log and is given by $1-1/\blocklength$.
This result follows from~\cite[Thm.~4.2]{lapidoth03-10a}, for the case $\blocklength=1$, and from~\cite[Eq.~(27)]{zheng02-02a}, for the case $\blocklength\geq 1$.
This concludes the proof.

For completeness, in \fref{lem:SIMO_rank_one} below we restate~\cite[Eq.~(27)]{zheng02-02a} for the SIMO case, and provide an alternative, much simpler proof of this result in \fref{sec:proof_of_lemma_rank_one}  below.
 \begin{lem}\label{lem:SIMO_rank_one}
 	The capacity of the SIMO channel~\eqref{eq:IO_relation} with $\RXant$ receive antennas, $\rankcorr=1$, and $\blocklength\geq 2$  is given by
 	\begin{align}\label{eq:capacity_rank_1_case}
 		\capacity(\snr)=  \left(1-{1}/{\blocklength}\right)\log\snr + \asconst, \quad\snr\rightarrow \infty.
 	\end{align}
 \end{lem}
%
%
%
%
% subsection second_part_ (end)

\subsection{Proof of \fref{lem:SIMO_rank_one}} % (fold)
\label{sec:proof_of_lemma_rank_one}

\subsubsection{Geometric Intuition} % (fold)
\label{sec:geometric_intuition}
When $\rankcorr = 1$, we can rewrite the IO relation as
\begin{equation*}
\matY=\vecs\tp{\vecx}+\matW
\end{equation*}
where $\vecs \sim\jpg(\veczero,\matI_{\RXant})$.
We next provide a geometric argument illustrating why the SIMO capacity pre-log coincides with the SISO capacity pre-log when $\rankcorr=1$.
A similar argument can be found in~\cite{zheng02-02a}.
Let $\vecx$ be an arbitrary vector in $\complexset ^\blocklength$.
In the absence of noise, the rows of \matY are collinear with $\vecx$.
The only information the receiver can recover (in the absence of noise) about the transmit vector \vecx from any of these rows is the line on which \vecx lies.
A line in $\complexset^{\blocklength}$ is characterized by $\blocklength-1$ complex parameters.
Hence, as argued in~\cite{durisi11-08a}, the receive signal $\matY$ carries $\blocklength-1$ parameters describing \vecx.
This number, divided by $\blocklength$,  coincides with the capacity pre-log we want to establish.
As one column of \matY is sufficient to recover the $\blocklength-1$ parameters describing the line on which \vecx lies, adding more receive antennas does not appear to be beneficial.
We next prove this result by sandwiching capacity between a lower bound and an upper bound that are  tight at high SNR.

\subsubsection{A Capacity Lower Bound} % (fold)
\label{sec:a_pre_log_lower_bound}
The RHS of~\eqref{eq:capacity_rank_1_case}  is a lower-bound on capacity.
This result follows directly from~\cite[Prop.~7]{liang04-09a}.

% subsubsection a_pre_log_lower_bound (end)

\subsubsection{A Matching Upper Bound Through Duality} % (fold)
\label{sec:duality_based_proof}
Establishing an asymptotically tight capacity upper bound is more involved.
Our proof is based on duality~\cite{lapidoth03-10a}, a technique that allows us to obtain a tight upper bound on $I(\vecx;\matY)$ by carefully choosing a probability distribution on $\matY$.
More precisely, let $\chtran(\cdot\given\vecx)$  denote the conditional distribution of $\matY$ given $\vecx$, and let $\mathsf{QW}$ denote the distribution induced on $\matY$ by the input distribution $\mathsf{Q}$ and by the channel $\chtran(\cdot\given\vecx)$.
Finally, let  $\mathsf{R}$ be an arbitrary distribution on $\matY$ with probability density function (pdf) $\mathsf{r}(\matY)$.
We use duality to upper-bound the mutual information $\mi(\vecx;\matY)$ as follows~\cite[Thm. 5.1]{lapidoth03-10a}:
\begin{align}
\label{eq:duality-expression}
\mi(\vecx;\matY) &\leq \Ex{\mathsf{Q}}{\mathrm{D}\lefto(\mathsf{W}(\cdot\given \vecx)\| \mathsf{R}(\cdot)\right) }\notag\\
&=-\Ex{\mathsf{QW}}{\log\mathsf{r}(\matY)}-\difent(\matY\given \vecx).
\end{align}
To get a tight capacity upper bound, the output distribution $\mathsf{R}$ must be chosen appropriately.
 For the SISO case, this choice can be motivated as follows: the geometry unveiled in \fref{sec:geometric_intuition} suggests to use the subspace spanned by $\vecx$ to convey information.
 This can be achieved by choosing an input distribution that is uniformly distributed on the sphere in $\complexset^{\blocklength}$ with radius $\sqrt{\blocklength\snr}$.
 The output distribution induced by this input distribution in the absence of additive noise turns out to yield a tight capacity upper bound, as shown in \cite{durisi11-08a}.

 Generalizing this approach to the SIMO case is not straightforward.
 The reason is as follows: for any choice of the input distribution, the matrix $\vecs\tp{\vecx}$ has rank at most~$1$, whereas the additive noise matrix \matW has full rank with probability one.
 This implies that, independently of the choice of the input distribution, the induced output distribution  in the absence of additive noise is not absolutely continuous~\cite[Def.~6.7]{rudin87a} with respect to $\mathsf{W}(\cdot\given \vecx)$, and, hence, the RHS of~\eqref{eq:duality-expression} diverges.
% But this implies that the RHS of (\markred{insert right equation number}) diverges.
% %
% %
To get a tight bound, one needs to choose an output distribution for which $\matY$ has full rank with probability one.
This implies that, differently from the SISO case, the additive noise needs to be accounted for in the choice of the output distribution.

To shed light on how this can be done, it is convenient   to express $\matY$ in terms of its singular-value decomposition (SVD).
More specifically, let $P = \min \{\RXant, \blocklength \}$ and $L = \max \{\RXant, \blocklength \}$; then~$\matY$ can be written  as  $\matY=\matU \mathbf{\Sigma} \herm{\matV}$, where $\matU \in \complexset^{\RXant\times P}$ and $\matV \in \complexset^{\blocklength \times P}$ are (truncated) unitary matrices, and $\mathbf{\Sigma}=\diag\{[\sigma_1(\matY)\, \cdots\, \sigma_{P}(\matY)]\}$ contains the  singular values of $\mathbf{Y}$ in descending order.
To make the SVD unique, we assume that the first row of $\matU$ is real and non-negative.
% Therefore, $\matU$ only occupy a submanifold $\tilde{\mathrm{S}}(\RXant,P)$ of the Stiefel manifold $\mathrm{S}(\RXant,P)$.
%
We shall take an output distribution for which $\sigma_1(\matY)$ is distributed as the nonzero singular value of the noiseless receive matrix $\vecs\tp{\vecx}$ and the remaining singular values are distributed as the ordered singular values of a $(\RXant-1) \times (\blocklength-1)$ random matrix with \iid $\jpg(0,1)$ entries.
More specifically, we take\footnote{We shall indicate $\sigma_i(\matY)$ simply as $\sigma_i$ whenever no ambiguity occurs.}
\begin{equation*}
\textsf{r}(\sigma_1,\cdots,\sigma_P) = \textsf{r}(\sigma_1)\cdot\textsf{r}(\sigma_2,\cdots,\sigma_P)
\end{equation*}
where
\begin{equation*}
%\label{eq:distribution-sigma1}
\textsf{r}(\sigma_1) = \frac{2\sigma_1}{\RXant\blocklength\rho} \cdot e^{-\sigma_1^2 / (\RXant\blocklength\rho)}, \quad \sigma_1>0
\end{equation*}
and~\cite[Thm.~2.17]{tulino04a}
\begin{multline*}
%\label{eq:distribution-sigma2M}
\textsf{r}(\sigma_2,\cdots,\sigma_P) = 2^{P-1}e^{-\sum\nolimits_{i=2}^P \sigma_i^2}\cdot \prod \limits_{i=2}^P \frac{\sigma_i^{2(L-P)+1}}{(L-i)!(P-i)!} \\ \cdot  \prod \limits_{i=2}^{P-1} \prod \limits_{j=i+1}^{P} \left( \sigma_i^2-\sigma_j^2\right)^2, \quad \sigma_2,\dots, \sigma_P>0.
\end{multline*}
Finally, we take $\matV$ and $\matU$ independent of the singular values and uniformly distributed (with respect to the Haar measure) on the Stiefel manifold\footnote{The set of complex $m\times n$ ($n\geq m$) unitary matrices form a manifold $\mathrm{S}(n,m)$ of $2mn-m^2$ real dimensions, called the Stiefel manifold~\cite{boothby86-a,zheng02-02a}.
This manifold has volume $\abs{\mathrm{S}(n,m)} = \prod_{i=n-m+1}^{n} {2\pi^i}/{(i-1)!}$.} $\mathrm{S}(\blocklength,P)$, and on the submanifold of $\mathrm{S}(\RXant,P)$  induced by the nonnegativity of the first row of $\matU$, respectively.
We next evaluate the RHS of~\eqref{eq:duality-expression} for the resulting output pdf, which we (still) denote by $\textsf{r}(\matY)$.
The conditional differential entropy $\difent(\matY\given \vecx)$ in \fref{eq:duality-expression}  can be easily computed:
\begin{equation}
\label{eq:h_y_x}
\difent(\matY \given \vecx) = \RXant \Ex{\vecx}{\log \left(\|\mathbf{x}\|^2+1\right)}+\RXant N\log(\pi e).
\end{equation}
% %
To evaluate the first term on the RHS of~\eqref{eq:duality-expression}, it is convenient to express $\mathsf{r}(\matY)$ in the SVD coordinate system.
By the change of variables theorem~\cite[Thm.~7.26]{rudin87a}, we get
\begin{multline}
\label{eq:change-variable}
-\Ex{\mathsf{QW}}{\log\mathsf{r}(\matY)} = -\Ex{\mathsf{QW}}{\log\mathsf{r}(\matU,\mathbf{\Sigma},\matV)} \\
+ \Ex{\mathsf{QW}}{\log J_{\RXant,\blocklength}(\sigma_1,\cdots,\sigma_{P})}
\end{multline}
where $J_{\RXant,\blocklength}(\sigma_1,\cdots,\sigma_{P})$ is the Jacobian of the SVD, which is given by \cite[App.~A]{zheng02-02a}
\begin{multline*}
J_{\RXant,\blocklength}(\sigma_1,\cdots\!,\sigma_{P})= \prod \limits_{i=1}^{P} \sigma_i^{2\left(L-P\right)+1} \cdot  \prod \limits_{i=1}^{P-1} \prod \limits_{j=i+1}^{P} \!\!\left( \sigma_i^2-\sigma_j^2\right)^2.
\end{multline*}
By construction, we have that
\begin{align}	
	\label{eq:entropy_term}
 &-\Ex{\mathsf{QW}}{\log\mathsf{r}(\matU,\mathbf{\Sigma},\matV)}=\underbrace{-\Ex{\mathsf{QW}}{\log\mathsf{r}(\matU)}-\Ex{\mathsf{QW}}{\log\mathsf{r}(\matV)}}_{=\landauO(1),\quad \snr \to \infty}	 \notag\\
	&\quad\quad-\Ex{\mathsf{QW}}{\log\mathsf{r}(\sigma_1)}-\Ex{\mathsf{QW}}{\log\mathsf{r}(\sigma_2,\dots,\sigma_P)}\notag\\
	   &=\log\snr-\Ex{\mathsf{QW}}{\log\sigma_1}\notag\\
		&\quad\quad + \underbrace{\Ex{\mathsf{QW}}{{\sigma_1^2}}/(\RXant\blocklength\snr)+\Ex{\mathsf{QW}}{\sum\limits_{i=2}^P\sigma_i^2}}_{\define c_1(\snr)}\notag \\
		&\quad\quad - \sum\limits_{i=2}^{P-1}\sum\limits_{j=i+1}^{P}\Ex{\mathsf{QW}}{\log (\sigma_i^2-\sigma_j^2)^2} \notag\\
		&\quad\quad - \sum\limits_{i=2}^{P} \Ex{\mathsf{QW}}{\log \sigma_i^{2(L-P)+1}}+\asconst,\quad \snr\to\infty.
\end{align}
The expectation of the Jacobian in~\fref{eq:change-variable} can be rewritten as
\begin{align}
\label{eq:Jacobian}
& \mathbb{E}_{\mathsf{QW}}\lefto[\log J_{\RXant,\blocklength}(\sigma_1,\cdots,\sigma_{P}) \right]\notag\\
& = \Ex{\mathsf{QW}}{\log \sigma_1^{2(L-P)+1}}  + \sum\limits_{j=2}^{P}\mathbb{E}_{\mathsf{QW}} \underbrace{\left[\log (\sigma_1^2-\sigma_j^2)^2\right] }_{\leq \log \sigma_1^4}\notag\\
&\quad + \sum\limits_{i=2}^{P} \Ex{\mathsf{QW}}{\log \sigma_i^{2(L-P)+1}} \notag \\
& \quad + \sum\limits_{i=2}^{P-1}\sum\limits_{j=i+1}^{P}\Ex{\mathsf{QW}}{\log (\sigma_i^2-\sigma_j^2)^2}.
\end{align}
Substituting \fref{eq:entropy_term} and \fref{eq:Jacobian} into \fref{eq:change-variable}, we obtain
\begin{multline}
\label{eq:h-y}
-\Ex{\mathsf{QW}}{\mathsf{r}(\matY)} \leq\log\snr + (\blocklength+\RXant -2) \Ex{\mathsf{QW}}{\log \sigma_1^2} \\
+c_1(\snr)+ \asconst,\quad\snr \rightarrow \infty.
\end{multline}
% %
Finally, substituting~\eqref{eq:h-y} and~\eqref{eq:h_y_x} into~\eqref{eq:duality-expression}, we get
% %
%
\begin{multline*}
%\label{eq:I_x_y_2}
I(\vecx;\matY) \leq \log\rho +(N- 2)\Ex{\mathsf{QW}}{\log \sigma_1^2} + c_1(\snr) \\+\RXant\underbrace{\left(\Ex{\mathsf{QW}}{\log \sigma_1^2}-\Ex{\vecx}{\log \left(\|\mathbf{x}\|^2+1\right)} \right)}_{\define c_2(\snr)} + \asconst, \quad \snr \rightarrow \infty.
\end{multline*}
We conclude the proof by showing that, $\Ex{\mathsf{QW}}{\log \sigma_1^2}\leq \log \snr + \landauO(1)$, $\snr \to \infty$ and that  $c_1(\snr)$ and $c_2(\snr)$ can be upper-bounded by finite constants.
For the first term, we have that
\begin{align}
	\label{eq:log_bound}
	\Ex{\mathsf{QW}}{\log \sigma_1^2}&\leq \Ex{\mathsf{QW}}{\log \tr\{\herm{\matY}\matY\}} \stackrel{(a)}{\leq} \log \sum_{i=1}^{\blocklength}\Ex{\mathsf{QW}}{\vecnorm{\vecy_i}^2}  \notag\\
	&\stackrel{(b)}{\leq} \log\snr+\landauO(1), \quad \snr \to \infty.
\end{align}
Here, in~(a) we used Jensen's inequality and (b) follows from~\eqref{eq:average-power-constraint}.
To show that $c_1(\snr)$ and $c_2(\snr)$ are bounded, the following lemma will turn out to be useful.
\begin{lem}[{\cite[Sec.~7.3]{horn85a}}]
\label{lem:lemma-sigma}
Let $\matA$, $\matB\in\complexset^{m\times n}$ and $p=\min\{m,n\}$.
Then
\begin{equation*}
%\label{eq:sigma-lamma}
\sigma_{i+j-1}(\matA+\matB)\leq \sigma_i(\matA)+\sigma_j(\matB),\quad 1\leq i,j\leq p,~\ i+j\leq p+1.
\end{equation*}
% In particular, \fref{eq:sigma-lamma} implies that $\sigma_1(\matA+\matB)\leq \sigma_1(\matA)+\sigma_1(\matB)$ and $\sigma_i(\matA+\matB)\leq \sigma_2(\matA)+\sigma_{i-1}(\matB)$, $i\geq 2$.
\end{lem}

If we choose $\matA=\vecs\tp{\vecx}$ and $\matB=\matW$, we obtain from \fref{lem:lemma-sigma}  that
\begin{equation}
\label{eq:sigma-all-bound}
\sigma_i(\matY) \leq
\begin{cases}
\|\vecs\|\|\vecx\|+\sigma_1(\matW), & i=1\\
\sigma_{i-1}(\matW), & 2\leq i \leq P.
\end{cases}
\end{equation}
By using~\eqref{eq:sigma-all-bound}, it follows that
\begin{align*}
\Ex{\mathsf{QW}}{\sum\limits_{i=2}^P \sigma_i^2(\matY)} \leq  \Ex{\matW}{\sum\limits_{i=1}^{P-1}  \sigma_i^2(\matW)}\leq \RXant\blocklength.
\end{align*}
This inequality, together with the inequality
\begin{equation*}	
	 \Ex{\mathsf{QW}}{\sigma_1^2}\leq \RXant\blocklength(\snr+1)
\end{equation*}
which can be established using similar steps to the ones leading to~\eqref{eq:log_bound}, are sufficient to   conclude  that $c_1(\snr)$ is bounded.

To establish that $c_2(\snr)$ is bounded, we start by noting that the first term in the expression that defines $c_2(\snr)$ can be upper-bounded as follows:
\begin{align*}
	 \Ex{\mathsf{QW}}{\log \sigma_1^2(\matY)} &\stackrel{(a)}{\leq} 2 \Ex{\mathsf{QW}}{\log\bigl(\|\vecs\|\|\vecx\|+\sigma_1(\matW)\bigr)} \notag \\
	&\stackrel{(b)}{\leq} 2 \Ex{\vecx}{\log\bigl(\Ex{\vecs,\matW} {\|\vecs\|\|\vecx\| +\sigma_1(\matW)}\bigr)} \\
	&\stackrel{(c)}{\leq} \Ex{\vecx}{\log\bigl(\sqrt{\RXant}(\vecnorm{\vecx}+\sqrt{\blocklength}) \bigr)^2}.
\end{align*}
Here,~(a) follows from~\eqref{eq:sigma-all-bound},~(b) holds because of Jensen's inequality, and in~(c) we used that $\Ex{}{\vecnorm{\vecs}}\leq \sqrt{\RXant}$ and   that
\begin{align*}
	\bigl(\Ex{}{\sigma_1(\matW)}\bigr)^2 &\leq     \Ex{}{\bigl(\sigma_1(\matW)\bigr)^2 }  \leq \Ex{}{\tr\bigl\{\herm{\matW}\matW\bigr\}} \\
	&=\RXant\blocklength.
\end{align*}
%
%\bigl(\Ex{}{\sigma_1(\matW)}\bigr)^2\leq \RXant\blocklength$.
%
Hence,
\begin{align*}
	c_2(\snr)&\leq \Ex{\vecx}{\log\frac{\bigl(\sqrt{\RXant}(\vecnorm{\vecx}+\sqrt{\blocklength})\bigr)^2}{\vecnorm{\vecx}^2+1}}  \\
	&\leq \sup_{\vecx} \lefto\{\log \frac{\bigl(\sqrt{\RXant}(\vecnorm{\vecx}+\sqrt{\blocklength})\bigr)^2}{\vecnorm{\vecx}^2+1}\right\} =\log [\RXant(\blocklength+1)].
\end{align*}
This concludes the proof.

% section the_simo_capacity_pre_log (end)
\bibliographystyle{IEEEtran}
\bibliography{IEEEabrv,publishers,confs-jrnls,giubib}

\end{document}